\documentclass[a4paper]{article}  

\usepackage{amsfonts}
\usepackage{amsmath} 
\usepackage{graphicx}
\usepackage{epstopdf}
\usepackage{multirow}
\graphicspath{{./Figures/}{./}}

\newcommand{\ch}[1]{{\color{black} #1}}   

\usepackage{epsfig} 
\usepackage{ctable}

\newcommand{\R}{{\mathbb R}}

\title{Information Theory to probe Intrapartum Fetal Heart Rate Dynamics}

\author{Carlos Granero-Belinchon$^{(1)}$, St\'ephane G. Roux$^{(1)}$, Patrice Abry$^{(1)}$, \\
Muriel Doret$^{(2)}$, Nicolas B. Garnier$^{(1)}$ \\
{\footnotesize $^{1}$ \quad Univ Lyon, Ens de Lyon, Univ Claude Bernard, CNRS, Laboratoire de Physique, F-69342 Lyon, France;} \\
  {\tt \footnotesize firstname.lastname@ens-lyon.fr}\\
{\footnotesize $^{2}$ \quad Femme-M\`ere-Enfant Hospital, Bron,, Univ. Lyon I, Lyon, France;}\\
  {\tt \footnotesize muriel.doret-dion@chu-lyon.fr}
  }

\begin{document}
\maketitle

\abstract{Intrapartum fetal heart rate (FHR) monitoring constitutes a reference tool in clinical practice to assess the baby health status and to detect fetal acidosis. 
It is usually analyzed by visual inspection grounded on FIGO criteria.  
Characterization of Intrapartum fetal heart rate temporal dynamics remains a challenging task and continuously receives academic research efforts. 
Complexity measures, often implemented with tools referred to as \emph{Approximate Entropy} (ApEn) 
or \emph{Sample Entropy} (SampEn), have regularly been reported as significant features for intrapartum FHR analysis. 
We explore 
how Information Theory, and especially {\em auto mutual information} (AMI), is connected to ApEn and SampEn and can be used to probe FHR dynamics. 
Applied to a large (\ch{1404} subjects) and documented database of FHR data, collected in a French academic hospital, it is shown that 
i) auto mutual information outperforms ApEn and SampEn for acidosis detection in the first stage of labor and continues to yield the best performance in the second stage; 
ii) Shannon entropy increases as labor progresses, and is always much larger in the second stage;
iii) babies suffering from fetal acidosis additionally show more structured temporal dynamics than healthy ones and that this progressive structuration can be used for early acidosis detection.
}

\section{Introduction}
\label{sec:intro}

\paragraph{\bf Intrapartum fetal heart rate monitoring} Because it is likely to provide obstetricians with significant information related to the health status of the fetus during delivery, 
intrapartum fetal heart rate (FHR) monitoring is a routine procedure in hospitals. 
Notably, it is expected to permit to detect fetal acidosis, which may induce severe consequences for both the baby and the mother and thus requires a timely and relevant decision for rapid intervention and operative delivery~\cite{Chandraharan2007}. 
In daily clinical practice, FHR is mostly inspected visually, trained from clinical guidelines formalized by the International Federation of Gynecology and Obstetrics
(FIGO)~\cite{Ayres-de-Campos2015,Figo1986}.
However, it has been well documented that such visual inspection is prone to severe inter-individual variability and even shows a substantial intra-individual variability \cite{Hruban2015jep}.
This is reflecting both that FHR temporal dynamics are complex and hard to assess, and that FIGO criteria lead to a demanding evaluation, as they mix several aspects of FHR dynamics (baseline drift, decelerations, accelerations, long and short-term variabilities). 
Difficulties in performing objective assessment of these criteria has led to a substantial number of unnecessary Caesarean sections~\cite{Alfirevic2006}.
This has triggered a large amounts of researches world wide aiming both to compute in a reproducible and objective way the FIGO criteria~\cite{Ayres-de-Campos2015}, and to devise new \emph{signal processing} inspired features to characterize FHR temporal dynamics (cf. \cite{Spilka2012,Haritopoulos2016} for reviews). \\

\paragraph{\bf Related works}  After the seminal contribution in the analysis of heart rate variability (HRV) in adults \cite{Akselrod1981}, spectrum estimation has been amongst the first signal processing tool that has been considered for computerized analysis of FHR, either constructed on models driven by characteristic time scales~\cite{Goncalves2006,vanLaar2008,Siira2013} or scale-free 
paradigm~\cite{Francis2002,Doret2011,Doret2015plos}). 
Further, aiming to explore temporal dynamics beyond the mere temporal \ch{correlations}, 
several variations of nonlinear analysis have been envisaged \ch{both for antepartum and  intrapartum FHR} \cite{Echeverria2004}, based, e.g., on multifractal analysis~\cite{Doret2011}, scattering transforms~\cite{Chudacek2014scatter}, phase-driven synchronous pattern averages~\cite{Magenes2000} 
or complexity and entropy measures~\cite{Pincus1992,Pincus1995,Lake2002}. 
Interested readers are referred to e.g.~\cite{Spilka2012, Haritopoulos2016} for overviews.
There has also been several attempts to combine features different in nature by doing multivariate classification using supervised machine learning strategies (cf. e.g., \cite{Georgieva2013,Spilka2012, Czabanski2012,Warrick2010,Spilka2016jbhi}).

Measures from Complexity Theory or Information Theory remain however amongst the most used tools to construct HRV characterization. 
They are defined independently from (deterministic) dynamical system or from (random) stochastic process frameworks.
The former led to standard references, both for adult and \ch{for antepartum and intrapartum fetal heart rate analysis}: \emph{Approximate Entropy} (ApEn)~\cite{Pincus1992, Dawes1992}, \emph{Sample Entropy} (SampEn)~\cite{Richman2000}, which can be regarded as practical approximations to Kolmogorov-Sinai or Eckmann-Ruelle complexity measures.
The stochastic process framework leads to the definitions of Shannon and R\'enyi entropies and entropy rates. 
\ch{Both worlds are connected by several relations, cf. e.g., \cite{Grunwald2003,Lake2006} for reviews.}
Implementations of ApEn and SampEn rely on \emph{correlation integral} based algorithm (CI) \cite{Pincus1992, G83} while that of Shannon entropy rates may instead benefit from \emph{$k$-nearest neighbor} ($k$-NN) algorithm \cite{K87}, which brings robustness and improved performance to entropy estimation \cite{Porta2013,Spilka2014,Xiong2017}. 

\paragraph{\bf Labor stages}  

Automated FHR analysis is complicated by the existence of two distinct stages during labor. 
The dilatation stage (stage 1) consists in progressive cervical dilatation and regular contractions. 
The active pushing stage (stage 2) is characterized by a fully dilated cervix and expulsive contractions.
The most common approaches in FHR analysis consist in not distinguishing stages and performing a global analysis~\cite{Costa2009a,Warrick2010} or in focusing on stage 1 only, as it is better documented and usually shows data with better quality, cf. e.g.~\cite{Georgieva2014,Spilka2016jbhi}. 
Whether or not temporal dynamics associated to each stage are different has not been intensively explored yet (see a contrario \cite{Spilka2014cinc,Lim2014}).
However, recently, some contributions start to conduct systematic comparisons \cite{Spilka2016BSI,C.Granero-Belinchon2017}. 

\paragraph{\bf Goals, contributions and outline} The present contribution remains in the category of works aiming to design new efficient features for FHR, here based on advanced information theoretic concepts. 
These new tools are applied to a high quality large (\ch{1404} subjects) and documented FHR database collected across years in an academic hospital in France, and described in Section~\ref{sec:DB}. 
The database is split into two datasets associated each with one stage of labor, which enables us first to assess and compare acidosis detection performance achieved by the proposed  features independently on each stage, and second to address differences in FHR temporal dynamics between the two stages. 
Reexamining formally the definitions of entropy rates in Information Theory, Section~\ref{sec:IT} first establishes that they can be split into two components: Shannon entropy that quantifies data static properties, and Auto mutual information (AMI) that characterizes temporal dynamics in data, combining both linear and nonlinear (or higher order) statistics. 
ApEn and SampEn, defined from Complexity Theory, are then explicitly related to entropy rates, and hence to AMI 
(cf. Section~\ref{sec:CT-IT}). 
Estimation procedures for Shannon entropy, entropy rate and AMI, based on \emph{$k$-nearest neighbor} ($k$-NN) algorithms \cite{K87}, are compared to those of ApEn and SampEn, constructed on \emph{Correlation Integral} algorithms \cite{Pincus1992, Costa2002, G83}. 
Acidosis detection performances are reported in Section~\ref{sec:res:performances}.
Results are discussed in terms of quality versus analysis window size, $k$-NN or correlation integral based procedures, and differences between stage 1 and stage 2. 
Further, a longitudinal study consisting of sliding analysis in overlapping windows across the delivery process permits to show that processes characterizing stage 1 and stage 2 are different (Section~\ref{sec:results:dynamical}).
%

\section{Datasets: intrapartum fetal heart rate times series and labour stages. }
\label{sec:DB}

\noindent {\bf Data collection.} Intrapartum fetal heart rate data were collected at the academic Femme-M\`ere-Enfant hospital, in Lyon, France, during daily routine
monitoring across the years 2000 to 2010.
They are recorded using STAN S21 or S31 devices with internal scalp electrodes at 12bit resolution, 500 Hz sampling rate (STAN, Neoventa Medical, Sweden).
Clinical information was provided by the obstetrician in charge, reporting delivery conditions as well as the health status of the baby, notably the
umbilical artery pH after delivery and the decision for intervention due to suspected acidosis \cite{Doret2011st}.

\noindent {\bf Datasets.} For the present study, subjects were included using criteria detailed in \cite{Doret2011st,Spilka2016jbhi}, leading to a total of \ch{1404} tracing, lasting from 30 minutes to several hours.
These criteria essentially aim to reject subjects with too low quality recording (e.g., too many missing data, too large gaps, too short recordings, \ldots).
\ch{As a result, for subjects in the database, the average fraction of data missing in the last 20 minutes before delivery is less than 5\%.}
The first goal of the present work is to assess the relevance of new information theoretics measures; their robustness to poor quality data is postponed for future work.

The measurement of pH, performed by blood test immediately after delivery is systematically documented and used as ground-truth: 
When pH $ \leq 7.05$, the newborn is considered has having suffered from acidosis, and referred to as acidotic (A, \ch{pH $ \leq 7.05$}). 
Conversely, when pH $ > 7.05$, the newborn is considered not having suffered from acidosis during delivery, and termed normal (N, \ch{pH $ >7.05$}).
\ch{In order to have a meaningful pH indication, we retain only subjects for which the time between end of recording and birth is lower or equal to 10 min.}

Following the discussion above on labor stages, subjects are split into two different datasets. 
Dataset I consists of subjects for which delivery took place \ch{after  a short stage 2 (less than 15 min) or during stage 1 (stage 2 was absent)},
It contains 913 normal and 26 acidotic  subjects. 
Dataset II gathers FHR for delivery that took place \ch{after more than 15 min of stage 2}.
It contains 450 normal and 15 acidotic subjects.

\noindent {\bf Beat-per-Minute time series and preprocessing.} For each subject, the collection device provides us with a digitalized list of RR-interarrivals $\Delta_k$ in ms.
In reference to common practice in FHR analysis, and for ease of comparisons amongst subjects, RR-interarrivals are converted in regularly sampled Beat-per-Minute times series, by linear interpolation of the samples $\{\ldots, 36000/\Delta_k, \ldots\}$.
The sampling frequency has been set to $f_s = 10$Hz as FHR do not contain any relevant information above $3$Hz.

\section{Methods}
\label{sec:methods}

\paragraph{Outline}

We describe in this section the five features that we use to analyze heart rate signals.
We propose to apply Information Theory, as defined by Shannon, to the analysis of cardiac signals. We do so by computing the Shannon entropy, the Shannon entropy rate, and the auto mutual information. The first section below is devoted to the definition of these quantities, which provide three features.
The second section reports the definitions of two features rooted in Complexity theory: Approximate Entropy (ApEn) and Sample Entropy (SampEn), which are classics in cardiac signal analysis. Although we use them in practice only as benchmarks, we devote a last section to their relation with the new features we propose.

Information Theory and Complexity theory only differ in the nature of the objects under study. 
Information Theory, on one hand, aims to analyze random processes and defines functionals of probability densities. Complexity theory, on the other hand, aims to analyze signals produced by dynamical systems 
and assumes the existence of ergodic probability measures to describe the density of trajectories in phase space, so that they can be manipulated as probability densities. In this spirit, we consider through all this paper, the signals to analyze as random processes, although they indeed originate from a dynamical system.

\paragraph{Assumptions}

For the sake of simplicity in the description of the features, and for practical use, we assume that signals are monovariate (unidimensional) and centered (zero mean) because the five features we use are independent of the first moment of the probability density function.
We also assume that signals are stationary. Although this may seem at first a very strong assumption, it is very reasonable when examining \ch{time windows smaller than the natural time scale of the evolution between stages 1 and 2}, as we discuss in section~\ref{sec:results:dynamical}, \ch{and larger than events such as contractions}.
Finally, we also assume that the signals contain $N$ points, sampled at a constant frequency. All estimates depends on $N$ via finite size effects. In the following, we do not mention this dependence explicitly in the notations, and only compare features computed over the same window size. 

\paragraph{Time-embedding}

Because we are interested in the dynamics of the signal, we use the delay-embedding procedure introduced by Takens~\cite{Takens1981} in the context of dynamical systems. Its goal is to include time-correlations into the statistics. We construct, by sampling the initial process $X$ every $\tau$ points in time, a $m$-dimensional time-embedded process $X^{(m,\tau)}$ as:
\begin{equation}
\textbf{x}_t^{(m,\tau)}=\left(x_t, \,x_{t-\tau}, \, \cdots, \, x_{t-(m-1)\tau}\right)\,.
\end{equation}

In practice, we have a finite number $N$ of points in the time-series, so there are $N-m+1$ well-defined embedded vectors.

\subsection{Information Theory features}
\label{sec:IT}

We now briefly recall definitions from Information Theory introduced by Shannon~\cite{Shannon1948}. This paradigm aims to describe processes in terms of information, and it can be applied to any experimental signals.

\subsubsection{Shannon and R\'enyi Entropies}
\label{sec:Ent}

Shannon entropy  $H(X^{(m,\tau)})$ of a $m$-dimensional embedded process $X^{(m,\tau)}$ is a functional of its joint probability density function $p\left(\textbf{x}_t^{(m,\tau)}\right)$~\cite{Shannon1948}:
\begin{equation}
H(X^{(m,\tau)}) 
=- \int_{\R^m} p\left( \textbf{x}_t^{(m,\tau)} \right)\log \left( p\left(\textbf{x}_t^{(m,\tau)}\right) \right) {\rm d}\textbf{x}_t^{(m,\tau)} \,.
\label{eq:def:Shannon}
\end{equation}
which does not depend on $t$, thanks to the stationarity of the signal. 
Shannon entropy measures the total information contained in the process $X^{(m,\tau)}$.
For embedding dimension $m=1$, it is independent of the sampling parameter $\tau$ and we write in the following 
$H(X^{(1,\tau)})=H(X)$. 

R\'enyi $q$-order entropy $R_q(X^{(m,\tau)})$ is defined as another functional of the probability density~\cite{Renyi1961}:
\begin{equation}
R_q(X^{(m,\tau)})=\frac{1}{1-q} \log\left(   \int_{\R^m} p^q\left(\textbf{x}_t^{(m,\tau)}\right) d\textbf{x}_t^{(m,\tau)} \right) \,.
\label{eq:def:Renyi}
\end{equation}
When $q \rightarrow 1$, the R\'enyi $q$-order entropy converges to the Shannon entropy.

\subsubsection{Entropy Rates} 

Shannon entropy rate is defined as:
\begin{equation}
h(X) = \lim_{m \rightarrow \infty} \frac{H(X^{(m,\tau)})}{m} \,,
\label{eq:def:rate1}
\end{equation}
which can be shown to be equivalent to:
\begin{equation}
h(X) = \lim_{m \rightarrow \infty} \left( H(X^{(m+1,\tau)} - H(X^{(m,\tau)} \right) \,.
\label{eq:def:rate2}
\end{equation}

We define the $m$-order Shannon entropy rate as measuring how much the Shannon entropy increases between two consecutive time-embedding dimensions $m$ and $m+1$:
\begin{eqnarray}
h^{(m,\tau)}(X) &=&  H(X^{(m+1,\tau)}) - H(X^{(m,\tau)}) \label{eq:def:h:diff}  \\
                &=&  H(X_{t+\tau}|X_{t}^{(m,\tau)}) \label{eq:def:h:cond} \,.
\end{eqnarray}
It quantifies the variation of the total information in the time-embedded process when the embedding dimension $m$ is increased by 1. Interpreting eq.(\ref{eq:def:h:diff}), it measures information in the $m+1$ coordinates of $X_{t}^{(m+1,\tau)}$ that is not contained in the $m$ coordinates of $X_{t}^{(m,\tau)}$. Following eq.(\ref{eq:def:h:cond}), it can also be interpreted as  the new information brought by the extra sample $X_{t+\tau}$ when the set of $m$ samples, $X_{t}^{(m,\tau)}$, is already known.

R\'enyi order-$q$ entropy rate and $m$-order R\'enyi order-$q$ entropy rate can be defined in the same way, replacing Shannon entropy $H$ by R\'enyi order-$q$ entropy $R_q$ in eq.(\ref{eq:def:rate1}) and eq.(\ref{eq:def:h:diff}), respectively. 
Nevertheless, it should be emphasized that R\'enyi order-$q$ entropy is lacking chain rule of conditional probabilities as soon as $q\neq 1$; therefore, eq.(\ref{eq:def:h:cond}) does not hold for R\'enyi order-$q$ entropy, unless $q=1$ (Shannon entropy).

\subsubsection{Mutual Information}  

The Mutual Information (MI) of two processes measures the information they share~\cite{Shannon1948}.
MI is the Kullback-Leibler divergence~\cite{Kullback1951} between the joint probability function and the product of the marginals which would express the joint probability function if the two processes were independent.
For time-embedded processes $X^{(m,\tau)}$ and $Y^{(p,\tau)}$, it reads:
\begin{eqnarray}
I(X^{(m,\tau)},Y^{(p,\tau)}) &=& I(\textbf{x}_{t'}^{(m,\tau)}, \textbf{y}_t^{(p,\tau)})= \nonumber \\
&=& \int_{\R^{m+p}} p(\textbf{x}_{t'}^{(m,\tau)},\textbf{y}_t^{(p,\tau)}) \log \left( \frac{p(\textbf{x}_{t'}^{(m,\tau)},\textbf{y}_t^{(p,\tau)})}{p(\textbf{x}_{t'}^{(m,\tau)})p(\textbf{y}_t^{(p,\tau)})} \right) {\rm d}\textbf{x}_{t'}^{(m,\tau)} {\rm d}\textbf{y}_t^{(p,\tau)} \,.
\label{eq:def:MInf}
\end{eqnarray}
Mutual information is symmetrical with respect to its two arguments.
If $X$ and $Y$ are stationary processes, the mutual information $I(X^{(m,\tau)},Y^{(p,\tau)})$ depends only on $t'-t$, the time difference between $X^{(m,\tau)}$ and $Y^{(p,\tau)}$.

\paragraph{Auto Mutual Information} 

If $Y=X$ and $t'-t=p\tau$, the MI measures the information shared by two consecutive chunks of the same process $X$, both sampled at $\tau$. 
This quantity is sometimes called ``information storage''~\cite{C.2014,Faes2015,Xiong2017}, and we refer to it as the Auto Mutual Information (AMI) of the process $X$:
\begin{equation}
I^{(m,p,\tau)}(X)=  I(\textbf{x}_{t+p\tau}^{(p, \tau)}, \textbf{x}_{t}^{(m, \tau)}) \,.
\label{eq:def:MI}
\end{equation}
Remarking that the concatenation of $\textbf{x}_{t+p\tau}^{(p,\tau)}$ and $\textbf{x}_{t}^{(m,\tau)}$ is nothing but the $(m+p)$-dimensional embedded vector
\begin{equation*}
\left(\textbf{x}^{(p,\tau)}_{t+p\tau},\textbf{x}_t^{(m,\tau)} \right)=\textbf{x}_t^{(m+p,\tau)} \,,
\end{equation*}
the AMI depends on the embedding dimensions $(m,p)$ and the sampling time $\tau$ only.
AMI of order $(m,p)$ measures the shared information between consecutive $m$-points and $p$-points dynamics, {\em i.e.}, by how much the uncertainty of future $p$-points dynamics is reduced if the previous $m$-points dynamics is known. 

Thanks to the symmetry of the MI with respect to its two arguments, and invoking the stationarity of $X$, the AMI is invariant when exchanging $m$ and $p$:
\begin{equation}
I^{(m,p,\tau)}(X)=  I(\textbf{x}_{t+p\tau}^{(p, \tau)}, \textbf{x}_{t}^{(m, \tau)}) 
= I(\textbf{x}_{t+m\tau}^{(m, \tau)}, \textbf{x}_{t}^{(p, \tau)})
= I^{(p,m,\tau)}(X) \,.
\label{eq:AMI:symmetry}
\end{equation}
We emphasize that the MI ---~and therefore AMI~--- are defined only for the Shannon entropy.
The expression of the R\'enyi order-$q$ mutual information is not unique as soon as $q\neq1$, and we do not consider it here.

\paragraph{\bf Special case $p=1$} 

If $p=1$, the AMI is directly related to the Shannon entropy rate of order $m$:
\begin{equation}
h^{(m,\tau)}(X) = H(X) - I^{(m,1,\tau)}(X) \,,
\label{eq:def:h:MI}
\end{equation}
or equivalently
\begin{equation}
I^{(m,1,\tau)}(X) = H(X) - h^{(m,\tau)}(X) \,.
\label{eq:def:MI:h}
\end{equation}
Interestingly, this splits the entropy rate $h^{(m,\tau)}(X)$ in two contributions. The first one is the total entropy $H(X)$ of the process which only depends on the one-point statistics, and so does not describe the dynamics of $X$.
The second term is the AMI $I^{(m,1,\tau)}(X)$ which depends on the dynamics of the process $X$, irrespective of its variance~\cite{GBelinchon2016}. 

\paragraph{\bf Special case of a process with Gaussian distribution} 
For illustration, if $X$ is a stationary Gaussian process, hence fully defined by its variance $\sigma^2$ and normalized correlation function $c(\tau)$, we have:
\begin{eqnarray}
H^{(p,\tau)}(X) &=& \frac{p}{2}\log\left(2 \pi e \sigma^{2}\right)+\frac{1}{2} \log\left(|\Sigma^{(p)}|\right) \\
I^{(m,p,\tau)}(X) &=& \frac{1}{2}\log \left(\frac{|\Sigma^{(m)}||\Sigma^{(p)}|}{|\Sigma^{(m+p)}|} \right),  
\end{eqnarray}
\noindent where $\Sigma^{(m)}$ is the $m\times m$ correlation matrix of the process $X$; $\Sigma_{i,j}= c(|i-j|\tau)$ and $|\Sigma^{(1)}|=1$.
For the particular case $p=1$, we have: 
\begin{eqnarray}
H(X) &=& \frac{1}{2}\log\left(2 \pi e \sigma^{2}\right) \label{eq:H:Gaussian} \\
I^{(m,1,\tau)}(X) &=& \frac{1}{2}\log\left(\frac{|\Sigma^{(m)}|}{|\Sigma^{(m+1)}|}\right)  \\
h^{(m,\tau)}(X) &=& \frac{1}{2}\log\left(2\pi e\sigma^{2}\right) - \frac{1}{2}\log\left(\frac{|\Sigma^{(m)}|}{|\Sigma^{(m+1)}|} \right) \,,
\end{eqnarray}
which clearly illustrates the decomposition of the entropy rate according to eq.(\ref{eq:def:h:MI}): the first term $H(X)$ depends only on the static (one-point) statistics (via $\sigma^2$) and the second term $I^{(m,1,\tau)}(X)$ depends on the temporal dynamics (and in this simple case only on the dynamics, via the auto-correlation function $c(\tau)$).

\subsection{Features from Complexity Theory}
\label{sec:CT}

In the 1960's, Kolmogorov and Sinai adapted Shannon's Information Theory to the study of dynamical systems. The divergence of trajectories starting from different but undistinguishable initial conditions can be pictured as creating uncertainty, so creating information. Kolmogorov Complexity (KC), also known as the Kolmogorov-Sinai entropy and noted $h_{\rm KS}(\rho)$ in the following, measures the mean rate of creation of information by a dynamical system with ergodic probability measure $\rho$. KC is constructed exactly as the Shannon entropy rate from Information Theory, using eq.(\ref{eq:def:rate1}) and the same functional form as in eq.(\ref{eq:def:Shannon}), but using the density $\rho$ of trajectories in phase space instead of the probability density $p$.
In the early 1980's, the Eckmann-Ruelle entropy $K_2(\rho)$~\cite{G83, E85} was introduced following the same steps but using the functional form of the R\'enyi order-2 entropy (equation (\ref{eq:def:Renyi})). The interest of $K_2$ relies in its easier and hence faster computation from experimental time series, which was at the time a challenging issue.

\paragraph{Kolmogorov-Sinai and Eckmann-Ruelle entropies}

The ergodic theory of chaos provides a powerful framework to estimate the density of trajectories in the phase space of a chaotic dynamical system~\cite{E85}. For an experimental or numerical signal, it amounts to assimilating the phase space average to the time average. Given a distance $d(.,.)$ ---~usually defined with the ${\cal{L}}^2$ or the $\cal{L}^\infty$ norm~--- in the $m$-dimensional embedded space, the local quantity
\begin{equation}
C^m_i(\epsilon) = \frac{{\rm number~of}~j ~{\rm ~such ~that} ~d\left({\bf x}^{(m,\tau)}_i,{\bf x}^{(m,\tau)}_j\right)\leq \epsilon}{N-m+1} \,,
\label{eq:def:local_CI}
\end{equation}
provides, up to a factor $\epsilon$, an estimate of the local density $\rho$ in the $m$-dimensional phase space around the point ${\bf x}^{(m,\tau)}_i$. The following averages:
\begin{eqnarray}
\Phi^m(\epsilon) &=& \frac{1}{N-m+1}\sum_{i=1}^{N-m+1} \ln C^m_i(\epsilon) \,, \label{eq:def:Phi} \\
C^m(\epsilon)&=& \frac{1}{N-m+1}\sum_{i=1}^{N-m+1} C^m_i(\epsilon) \,, \label{eq:def:corr_int}
\end{eqnarray}
are then used to provide the following equivalent definitions of the complexity measures~\cite{E85}:
\begin{eqnarray}
h_{\rm KS}(\rho) &=& \lim_{\epsilon\rightarrow 0} \lim_{m\rightarrow \infty} \lim_{N\rightarrow \infty} 
\left( \Phi^m(\epsilon) - \Phi^{m+1}(\epsilon) \right) \,, 
\label{eq:def:KS:practical} \\
K_2(\rho) &=& \lim_{\epsilon\rightarrow 0} \lim_{m\rightarrow \infty} \lim_{N\rightarrow \infty} 
\ln\left( \frac{C^m(\epsilon)}{C^{m+1}(\epsilon)}\right) \,.
\label{eq:def:ER:practical} 
\end{eqnarray}

\subsubsection{Approximate Entropy}

Approximate Entropy (ApEn) was introduced by Pincus in 1991 for the analysis of babies heart rate~\cite{P91}. It is obtained by relaxing the definition (\ref{eq:def:KS:practical}) of $h_{\rm KS}$ and working with a fixed embedding dimension $m$ and a fixed box size $\epsilon$, often expressed in units of the standard deviation $\sigma$ of the signal as $\epsilon=r \sigma$. ApEn is defined as
\begin{equation}
{\rm ApEn}(m, \epsilon) = \Phi^m(\epsilon) - \Phi^{m+1}(\epsilon) \,.
\label{eq:ApEn:def}
\end{equation}
On the practical side, and in order to have a well-defined $\Phi^m(\epsilon)$ in (\ref{eq:def:Phi}), the counting of neighbors in the definition (\ref{eq:def:local_CI}) allows self-matches $j=i$. This ensures that $C^m_i(\epsilon)>0$, which is required by (\ref{eq:def:Phi}).
ApEn depends on the number of points $N$ in the time series. Assuming $N$ is large enough, we have 
\begin{equation}
\lim_{\epsilon\rightarrow 0} \lim_{m\rightarrow \infty} {\rm ApEn}(m,\epsilon) = h_{\rm KS}\,,
\label{eq:h_KS}
\end{equation}
We interpret ApEn as an estimate of the $m$-order Kolmogorov-Sinai entropy $h_{\rm KS}$ at finite resolution $\epsilon$.
The larger $N$, the better the estimate. 
More interesting is that the non-vanishing value of $\epsilon$ in its definition makes ApEn insensitive to details at scales lower than $\epsilon$. On one hand, this is very interesting when considering an experimental (therefore noisy) signal: choosing $\epsilon$ larger than the rms of the noise (if known) filters the noise, and ApEn is then expected to measure only the complexity of the underlying dynamics. This was the main motivation of Pincus and explains the success of ApEn. On the other hand, not taking the limits $\epsilon\rightarrow 0$ and $m\rightarrow \infty$ makes ApEn an ill-defined quantity that has no reason to behave like $h_{\rm KS}$. In addition, only very few analytical results have been reported on the bias and the variance of ApEn.

Although $m$ should in theory be larger than the dimension of the underlying dynamical system, ApEN is defined and used for any possible value of $m$ and most applications reported in the literature are using small $m$ (1 or 2) without any analytical support, but with great success~\cite{P91,Krstacic2002}.

\subsubsection{Sample Entropy}

A decade after Pincus seminal paper, Richman and Moorman pointed out that ApEn contains in its very definition a source of bias and was lacking in some cases "relative consistency". They defined Sample Entropy (SampEn) on the same grounds as ApEn:
\begin{equation}
{\rm SampEn}(m,\epsilon) = \ln\left( \frac{C^m(\epsilon)}{C^{m+1}(\epsilon)}\right) \,.
\label{eq:SampEn:def}
\end{equation}
so that
\begin{equation}
\lim_{\epsilon\rightarrow 0} \lim_{m\rightarrow \infty} {\rm SampEn} = K_2 \,,
\label{eq:K_2}
\end{equation}
On the practical side, the counting of neighbors in (\ref{eq:def:local_CI}) does not allow self-matches.
$C^m_i(\epsilon)$ may vanish, but when averaging over all points in eq.(\ref{eq:def:corr_int}), the correlation integral $C^m(\epsilon)>0$. In practice, SampEn is considered to improve on  ApEn as it shows 
lower sensitivity to parameter tuning and data sample size than ApEn \cite{R00,Lake:2011}. 

We interpret SampEn as an estimate of the $m$-order Eckmann-Ruelle entropy $K_2$ at finite resolution $\epsilon$.

\subsubsection{Estimation}
\label{sec:CT:estimation}

We note in the following ApEn$^{(m)}$ and SampEn$^{(m)}$ the estimated values 
of ApEn and SampEn using our own Matlab implementation, based on Physio-Net packages. 
We used the commonly accepted value, $\epsilon=0.2 \sigma$, with $\sigma$ the standard deviation of $X$, and $m=2$. 
For all quantities, we used $\tau = 5 = f_s/f_{\rm max}$ with $f_{\rm max} = 2 $Hz the cutoff frequency above which FHR times series essentially contain no relevant information~\cite{Doret2011}; this time delay corresponds to $0.5$s.

\subsection{Connecting Complexity Theory and Information Theory.}  
\label{sec:CT-IT}

We consider here for clarity only the relation between ApEn and $m$-order Shannon entropy rate, although the very same relation holds between SampEn and $m$-order R\'enyi order-2 entropy rate.
In Information theory terms, ApEn appears as a particular estimator of the $m$-order Shannon entropy rate 
that computes the probability density by counting, in the $m$-dimensional embedded space, the number of neighbors in a hypersphere of radius $\epsilon$, which can be interpreted as a particular kernel estimation of the probability density.

\subsubsection{Limit of large datasets and vanishing $\epsilon$: exact relation}

When the size $\epsilon$ of the spheres tends to 0, the expected value of ApEn for a stochastic signal $X$ with any smooth probability density is related, in the limit $N\rightarrow\infty$, 
to the $m$-order Shannon entropy rate~\cite{Lake2006}:
\begin{equation}
{\rm ApEn}(m,\epsilon) \underset{\epsilon\to 0}{\simeq}  h^{(m,\tau)}(X) - \log(2\epsilon)  \,.  \\
\end{equation}
Both terms involve $m$-points correlations of the process $X$.
This relation allows a quantitative comparison of ApEn with the $m$-order Shannon entropy rate $h^{(m,\tau)}$.
The $\log(2\epsilon)$ difference corresponds to the paving ---~with hyperspheres of radius $\epsilon$~--- of the continuous $m$-dimensional space over which the probability $p(\textbf{x}_t^{(m,\tau)})$ 
involved in eq.(\ref{eq:def:Shannon}) is defined, and thus $h^{(m,\tau)}$. This paving defines a discrete phase space, over which eqs. (\ref{eq:def:local_CI}), (\ref{eq:def:Phi}) and (\ref{eq:ApEn:def}) 
operate to define ApEn~\cite{Jaynes1962}.
This illustrates that, for a stochastic signal, ApEn diverges logarithmically as the size $\epsilon$ approaches 0, as expected for $h_{\rm KS}$.
Fortunately, $\epsilon$ is fixed in the definition of ApEn which allows in practice to compute it for any signal/process. 

\subsubsection{New features} 

Once recognized the success of ApEn and remembering its relation to $h^{(m,\tau)}$, it seems interesting to probe other $m$-order Shannon entropy rate estimators.
A straightforward improvement would be to consider a smooth ---~e.g. Gaussian~--- kernel of width $\epsilon$ instead of the step function used in (\ref{eq:def:local_CI}). We prefer to reverse the 
perspective and use a $k$-nearest-neighbor (\ch{$k$-NN}) estimate. Instead of counting the number of neighbors in a sphere of size $\epsilon$, this approach searches for the size of the sphere that accomodates $k$ neighbors. 
In practice, we compute the entropy $H$ with the Kozachenko \& Leonenko estimator~\cite{K87, Singh2003}, which we note $\hat{H}$. We compute the auto mutual information $I^{(m,p,\tau)}$ with the Kraskov {\em et al} estimator~\cite{Kraskov2004}, which we note $\hat{I}^{(m,p)}$. We then combine the two according to eq.(\ref{eq:def:h:MI}) to get an estimator $\hat{h}^{(m)}$ of the $m$-order Shannon entropy rate. 
We use $k=5$ neighbors and set $\tau=5$ (see section~\ref{sec:CT:estimation}).

We report in the next section our results for the five features when setting $m=2, p=1$, and compare their performances in detecting acidosis. 
The dependance of the $m$-order entropy rate (and its estimators) on $m$ is expected to give some insight in the dimension of the attractor of the underlying dynamical system, but as we have pointed out, the dynamics is indeeed contained in the AMI part of the entropy rate. This is why we further explore the effect of varying the embedding dimensions $m$ and $p$ on the AMI estimator $\hat{I}^{(m,p)}$.

\section{Results: Acidosis detection performance}
\label{sec:Res}
\label{sec:result}

\subsection{Comparison of features performance, using a single time window, just before delivery}
\label{sec:res:performances}

\paragraph{Average features value for normal and abnormal subjects} 

\begin{figure}
\includegraphics[width=\textwidth]{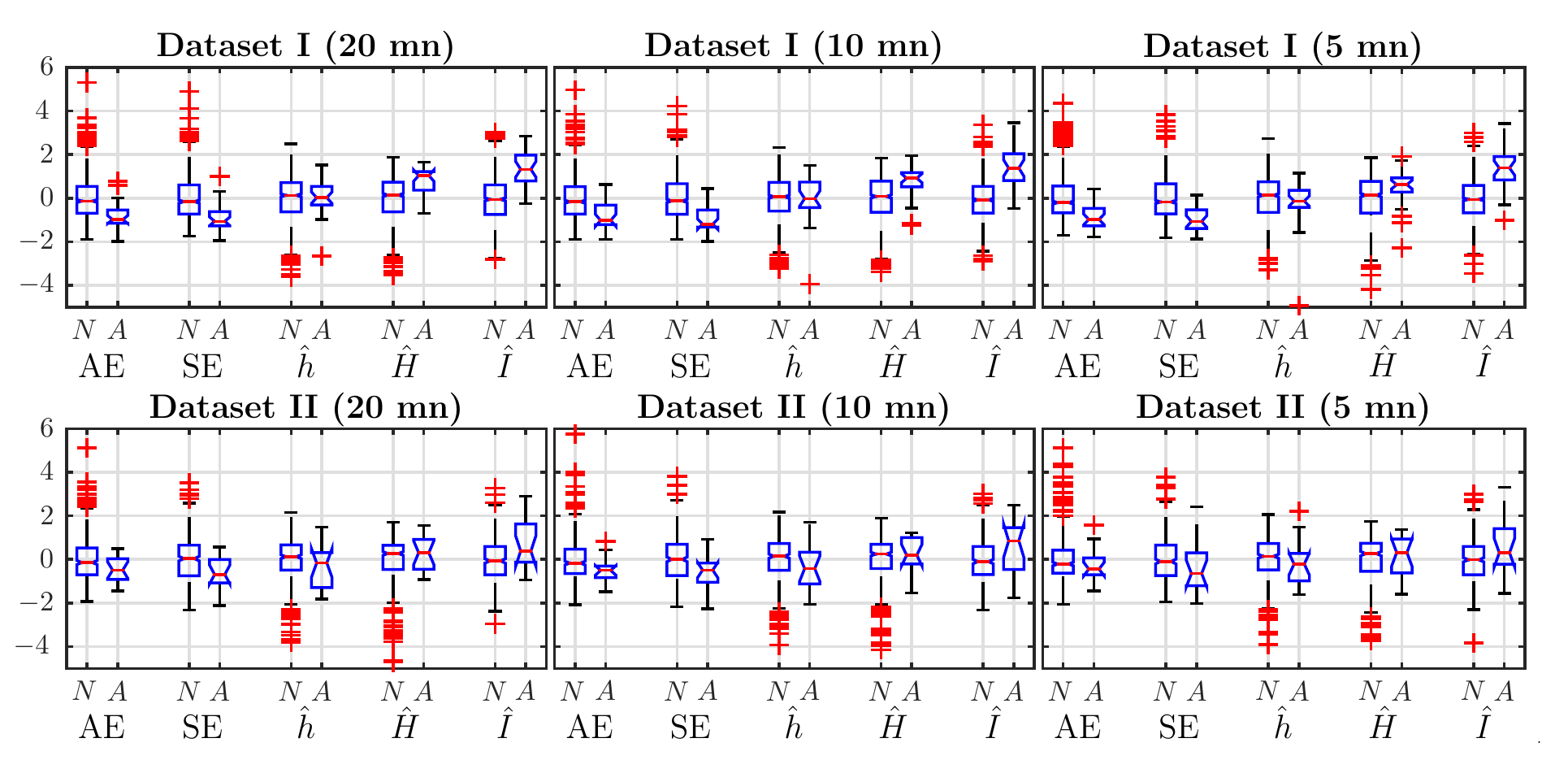}
\caption{Box plot based comparisons of the five different (normalized) estimates \ch{(ApEn$^{(m)}$ (AE), SampEn$^{(m)}$ (SE), Shannon entropy rate $h^{(m)}$ ($\hat{h}$), Shannon entropy ($\hat{H}$) and AMI $I^{(m,p)}$ ($\hat{I}$)}) for normal (N) and pathological (A for "abnormal") subjects, for dataset I (top) and dataset II (bottom). Each column corresponds to a different window size: $T=$ 20mn, 10mn and 5mn. All features are computed with $m=2$, $p=1$.}
\label{fig:boxplot20105}
\end{figure}

We compute the five features: ApEn$^{(m)}$, SampEn$^{(m)}$, $\hat{h}^{(m)}$, $\hat{H}$ and $\hat{I}^{(m,p)}$ for normal and acidotic (abnormal) subjects in dataset I and II using data from the last $T=20$mn before delivery, which are the most crucial. 
We use the classical values $m=2$ and $p=1$ for embedding dimensions.
To compare performance, we present the box plots of the five normalized (zero-mean, unit-variance) estimates in the left column of Fig.~\ref{fig:boxplot20105}.
For dataset I, the average of ApEn and SampEn for acidotic subjects is smaller than for normal subjects, while the average of Shanon entropy rate does not show any tendency. This is surprising as one might have expected for $\hat{h}^{(m)}$ a behavior similar to ApEn and SampEn (see section \ref{sec:CT-IT}).
Average values of $\hat{H}$ and $\hat{I}^{(2,1)}$ are larger for acidotic subjects.
The larger value of Shannon entropy $H$ indicates that the acidotic FHR signals contain more information. The larger value of AMI indicates \ch{a stronger dependence structure in the dynamics of} abnormal subjects.

For subjects in dataset II, it is harder to find any tendency by looking the average values.

\paragraph{Features performance}

Fetal acidosis detection performance is assessed with the $p$-value given by the classical Wilcoxon ranksum test.
This non-parametric test of the null hypothesis ---~which corresponds to identical medians of the distributions of estimates in the normal and abnormal classes~--- is reported in table~\ref{table1}.  
\begin{table}
\begin{center}
\begin{tabular}{||c||c|c||c|c||} \hline
 & \multicolumn{2}{|c||}{Dataset I} & \multicolumn{2}{c||}{Dataset II}  \\ \hline 
& AUC & $p$-value & AUC & $p$-value  \\ \hline 
 ApEn$^{(2)}$ & 0.76 & 4.08e-06 $\star\star$ & 0.61 & 1.33e-01  \\ \hline 
 SampEn$^{(2)}$ & 0.79 & 5.92e-07 $\star\star$ & 0.67 & 2.35e-02 $\star$  \\ \hline 
 $\hat{h}^{(2)}$ & 0.50 & 9.75e-01 & 0.39 & 1.36e-01  \\ \hline 
 $\hat{H}$ & 0.76 & 8.36e-06 $\star\star$ & 0.56 & 4.23e-01  \\ \hline 
 $\hat{I}^{(2,1)}$ & {\bf 0.84} & 2.00e-09 $\star\star$  & {\bf 0.68} & 1.69e-02 $\star$  \\ \hline
\end{tabular}
\end{center}
\caption{Area under \ch{Receiver Operational Characteristics} curves (as shown in Fig.~\ref{fig:ROC}) and $p$-value obtained from the Wilcoxon rank-sum test, for each of the five different estimates, all with $m=2$. }
\label{table1}
\end{table}
We have added one $\star$ symbol when the $p$-value is less than 0.05, two $\star\star$ when less than 0.01. We see that for dataset I, ApEn$^{(m)}$, SampEn$^{(m)}$, $\hat{H}$ and $\hat{I}^{(m,p)}$ for $m=2$ discriminate normal and acidotic subjects, while $\hat{h}^{(m)}$ does not.
Out of the three estimates (ApEn$^{(m)}$, SampEn$^{(m)}$, $\hat{h}^{(m)}$) based on entropy rates, the nearest-neighbors one for Shannon entropy rate is the poorest, although its decomposition into Shannon entropy 
(static one-point information) and AMI (which includes dynamic information) leads to two satisfying estimates
Fig.~\ref{fig:boxplot20105} and Table~\ref{table1} both show that the best performing estimators are SampEn$^{(m)}$ and $\hat{I}^{(m,p)}$. In dataset II, although all features performs more poorly than in dataset I, SampEn and AMI are again the best ones, with a $p$-value lower than 0.05. We focus on these two features in the following.

\paragraph{Receiver Operational Characteristics}

\begin{figure}
\begin{center}
\includegraphics[width=0.75\textwidth]{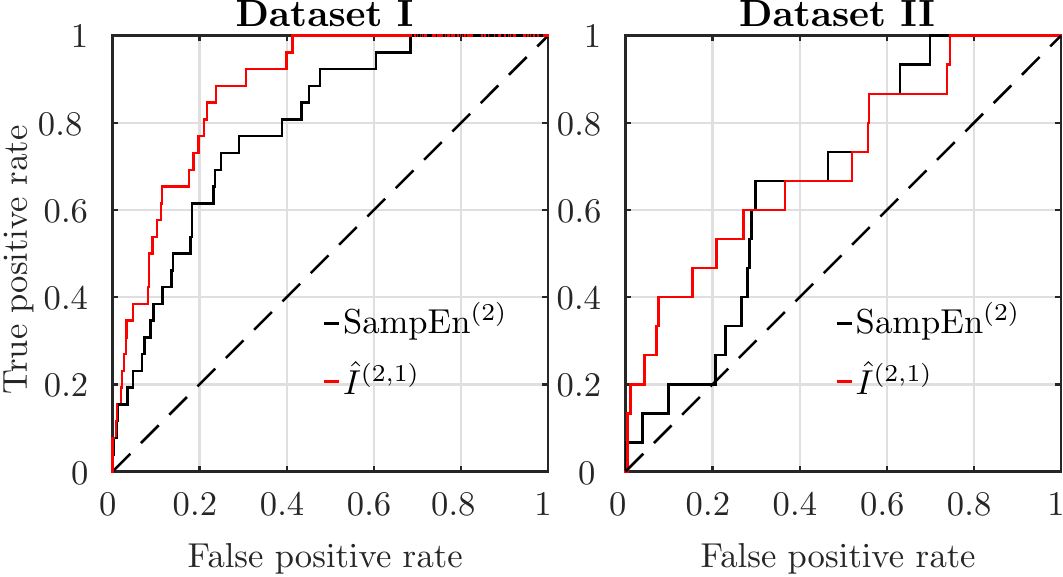}
\end{center}
\caption{\ch{Receiver Operational Characteristics} (ROC) curves for SampEn$^{(m)}$ (black) and AMI \ch{estimator $\hat{I}^{(m,p)}$} (red), for subjects in dataset I (left) and II (right). $m=2$, $p=1$.}
\label{fig:ROC}
\end{figure}

To compare the two best performing features, SampEn and AMI, we plot receiver operational characteristics (ROC) curves in Fig.~\ref{fig:ROC}, both for dataset I and II, using data from the last $T=20$mn before delivery.
For dataset I, AMI better discriminates acidotic subjects from normal ones. For dataset II, AMI discrimination is only slightly better than SampEn.
The area under the curve (AUC) of the ROC of the features is reported in Table~\ref{table1}, with a bold font to indicate the estimator with the largest AUC.
Performance is much worse in dataset II than in dataset I: the AUC is reduced. Nevertheless, AMI is always the better performing estimator (its AUC reduces from $0.84$ to $0.68$), followed by SampEn (AUC reducing from $0.79$ to $0.67$).

\subsection {Effect of the window size on the performance} 
We investigate the robustness of the detection performance when the window size $T$ is reduced, using data from $T=20,10$ and $5$ minutes. Results are reported in Fig~\ref{fig:boxplot20105} and Table~\ref{table2}. 
\begin{table}
\begin{center}
\begin{tabular}{||c|c||c|c||c|c||c|c||c|c||} \hline\hline 
 & & \multicolumn{2}{|c||}{SampEn$^{(2)}$} & \multicolumn{2}{c||}{$\hat{I}^{(2,1)}$} & \multicolumn{2}{c||}{$\hat{I}^{(3,3)}$}   \\ \hline 
& & AUC & $p$-value & AUC & $p$-value & AUC & $p$-value  \\ \hline \hline 
 \multirow{3}{*}{Dataset I}& 20 mn & 0.79 & 5.92e-07 $\star\star$ & 0.84 & 2.00e-09 $\star\star$ & {\bf 0.88} & 5.46e-11 $\star\star$  \\ 
 \cline{2-8}  & 10 mn& 0.76 & 1.22e-07 $\star\star$ &0.84 &2.22e-09 $\star\star$ & {\bf 0.87} & 1.40e-10 $\star\star$ \\  
 \cline{2-8} & 5 mn& 0.72 & 1.97e-07 $\star\star$ &0.83 &7.47e-09 $\star\star$  & {\bf 0.86} & 6.26e-10 $\star\star$ \\  
 \hline \hline 
 \multirow{3}{*}{Dataset II}& 20 mn & 0.67 & 2.35e-02 $\star$ & 0.68 & 1.69e-02 $\star$ & {\bf 0.71} & 5.36e-03 $\star\star$ 
 \\ \cline{2-8}  & 10 mn& 0.62 & 1.56e-02 $\star$ &0.64 &5.87e-02 & {\bf 0.68} & 1.66e-02 $\star$ \\  
 \cline{2-8} & 5 mn& 0.62 & 5.16e-02 &0.60 &1.70e-01 & {\bf 0.64} & 7.29e-02 \\  
 \hline \hline 
\end{tabular}
\caption{AUC and $p$-value of Wilcoxon test of SampEn and AMI in datasets I and II using data from the last 20, 10 or 5mn before delivery.}
\label{table2}
\end{center}
\end{table}
$p$-values and AUC both indicate that $\hat{I}^{(2,1)}$ and SampEn$^{(2)}$ provide robust discrimination in dataset I even when the observation length is reduced.
Again, $\hat{I}^{(2,1)}$ is better performing: its AUC is reduced from 0.84 to 0.83 when $T$ is reduced from 20mn to 5mn, where the AUC of SampEn is reduced from 0.79 to 0.72.
In dataset II, once again, performance degrades, but AMI is still better at discriminating acidotic from normal subjects.
In the following, we focus on AMI estimates only. 

\subsection{Effect of the embedding dimensions on (fetal acidosis detection) performance of AMI}

\begin{figure}
\includegraphics[width=0.9\textwidth]{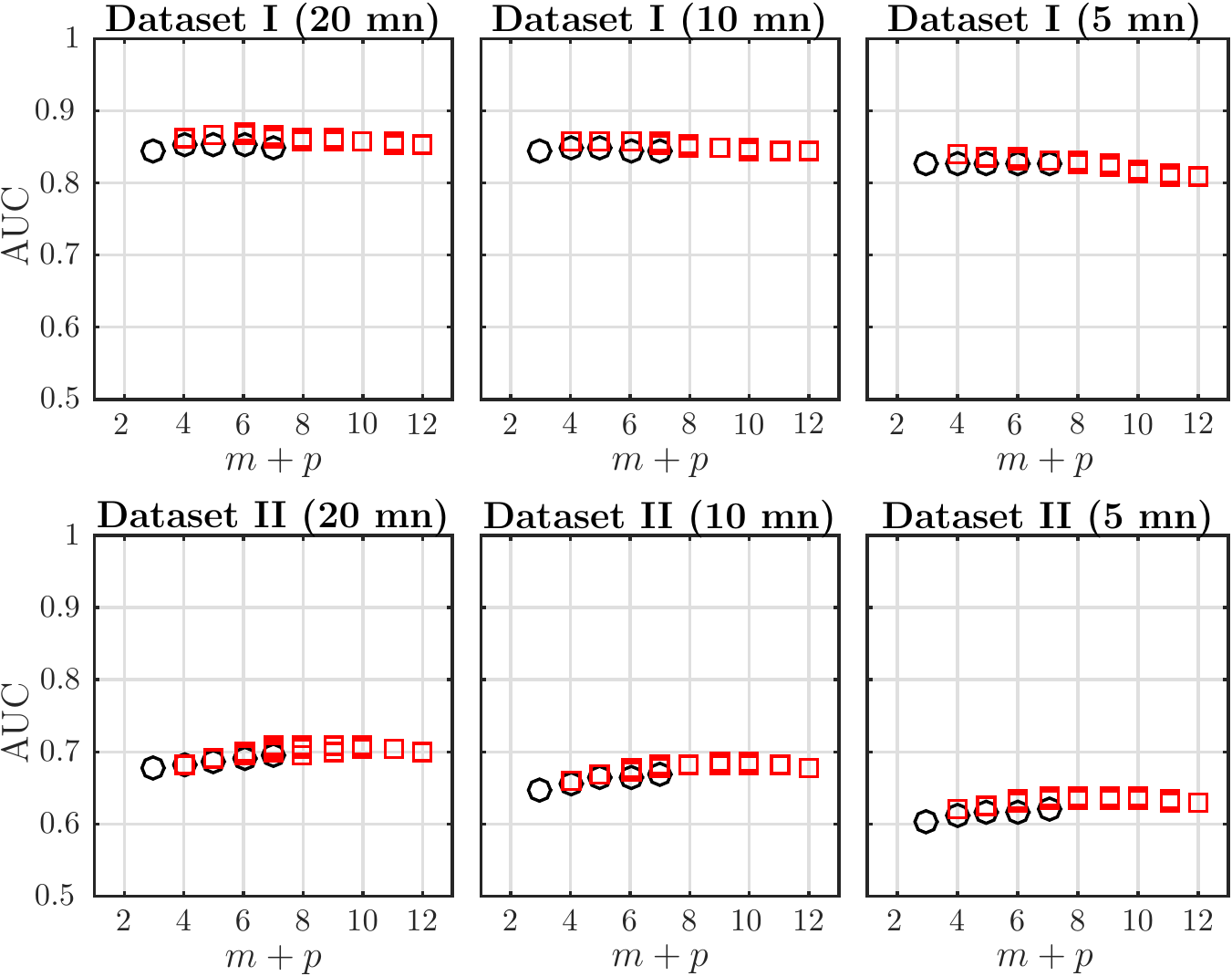} 
\caption{AUC of ROC for $\hat{I}^{(m,p)}$ as a function of the total embedding dimension $m+p$, with $m \ge p$, for time windows of size $T=20, 10, 5$mn, for data in dataset I (first line) and dataset II (second line).
Black circles indicate the special case $p=1$ corresponding to the classical definition of Shannon entropy rate (see eq. (\ref{eq:def:h:MI})).  
Red squares correspond to $p \ge 2$.}
\label{fig:aucmp}
\end{figure}

In order to improve the acidosis detection performance of the AMI, especially in dataset II, we increase the embedding dimensions $m$ and $p$ used in computing $\hat{I}^{(m,p)}$. This way, we probe higher order \ch{dependence stucture} in the dynamics. 
Because of the symmetry of AMI (eq.~(\ref{eq:AMI:symmetry})), \ch{and aiming at probing the effect of increasing either $m$ or $p$,} we plot the AUC of ROC as a function of $m+p$ \ch{only}. \ch{The dependence of AMI on $m-p$ is much smaller and not reported here.}
These computations have been done with a larger value $k=15$ in the \ch{$k$-NN} algorithms, in order to accommodate the possibly large embedding dimensions ($m+p$ up to 12).
Results are presented in Fig.~\ref{fig:aucmp}.

For a fixed window size, the AUC increases when $m+p$ increases, and reaches a maximum; it then remains constant or decreases slightly. This behaviour is observed in both datasets I and II and for any window size $T\in[5,10,20]$mn. 
Varying $T$ dos not seems to change the location of the maximum of the AUC in a given dataset.
The optimal embedding dimension is $m+p=6$ in dataset I and $m+p=10$ in dataset II. This hints at a difference in the dynamics of the FHR in the two datasets.
Because both bias and computation time increase with the total dimensionality~\cite{Gao2016}, the maximal embedding is restricted to $m=p=3$.
A reduction of the AUC is observed when the analysis window is reduced, but this is only significant for dataset II.
 
We reported in Table~\ref{table2} the AUC and \textit{p}-values of the AMI for two embedding dimensions:
$\hat{I}^{(2,1)}$ and $\hat{I}^{(3,3)}$, for datasets I and II and several window sizes. 
The best performing estimator is indicated in bold.
For all observation windows and for the two datasets, $\hat{I}^{(3,3)}$ achieves the best performance.
Their AUC is always larger than the one obtained using SampEn or AMI with $m=2$ and $p=1$. 

\subsection{Dynamical Analysis}
\label{sec:results:dynamical}

We now explore how long before delivery can the AMI diagnose fetal acidosis on a FHR signal.
To do so, we do not restrict our analysis to the last data points before delivery, but we apply it to an ensemble of windows scanning first and second stages of labor. We examine the dynamics of $\hat{I}^{(3,3)}$, the best performing feature, for both normal and abnormal subjects.

\subsubsection{Dataset I: rapid delivery}

\begin{figure}
\begin{center}
\includegraphics[width=0.45\textwidth]{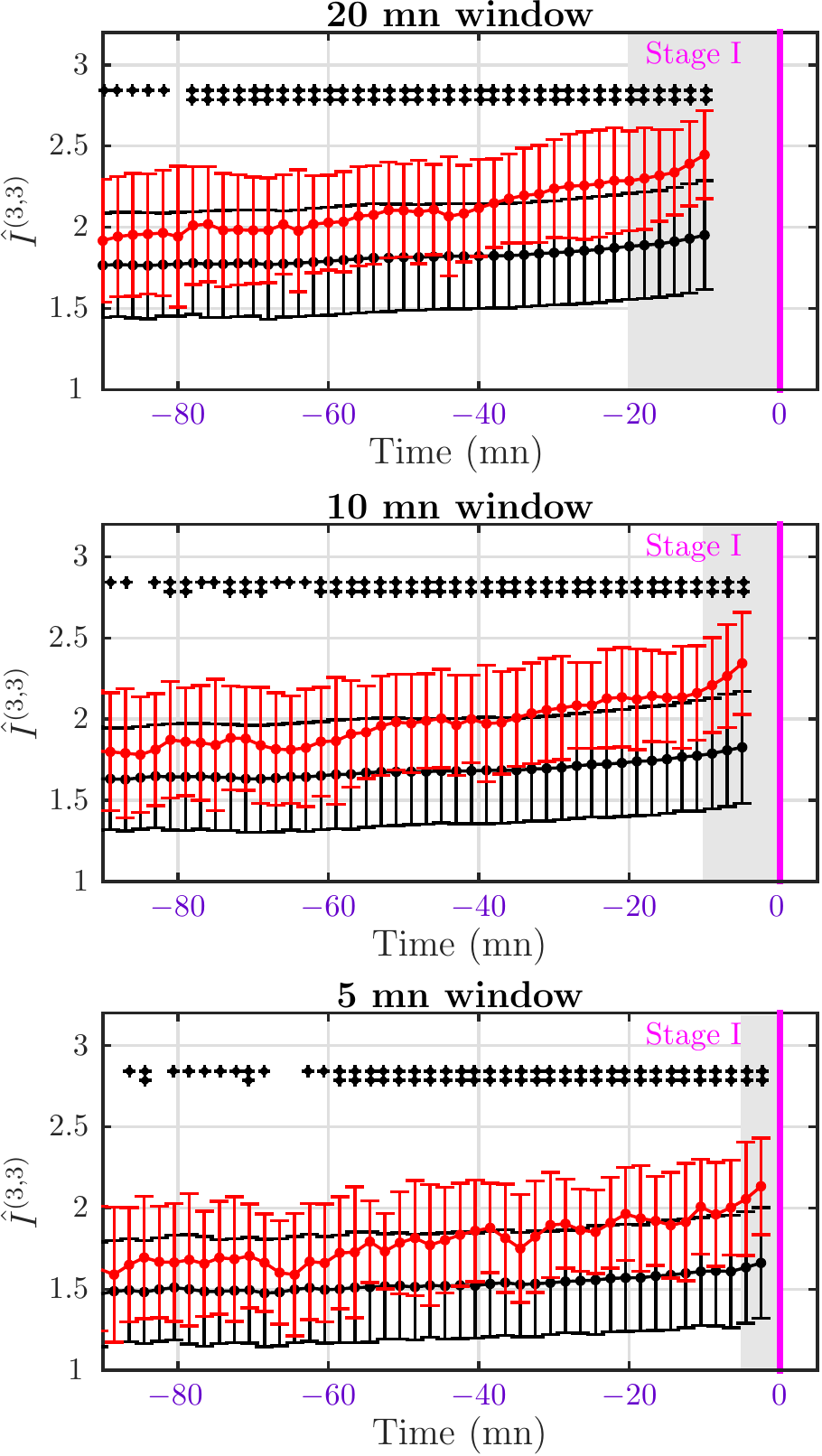}
\includegraphics[width=0.45\textwidth]{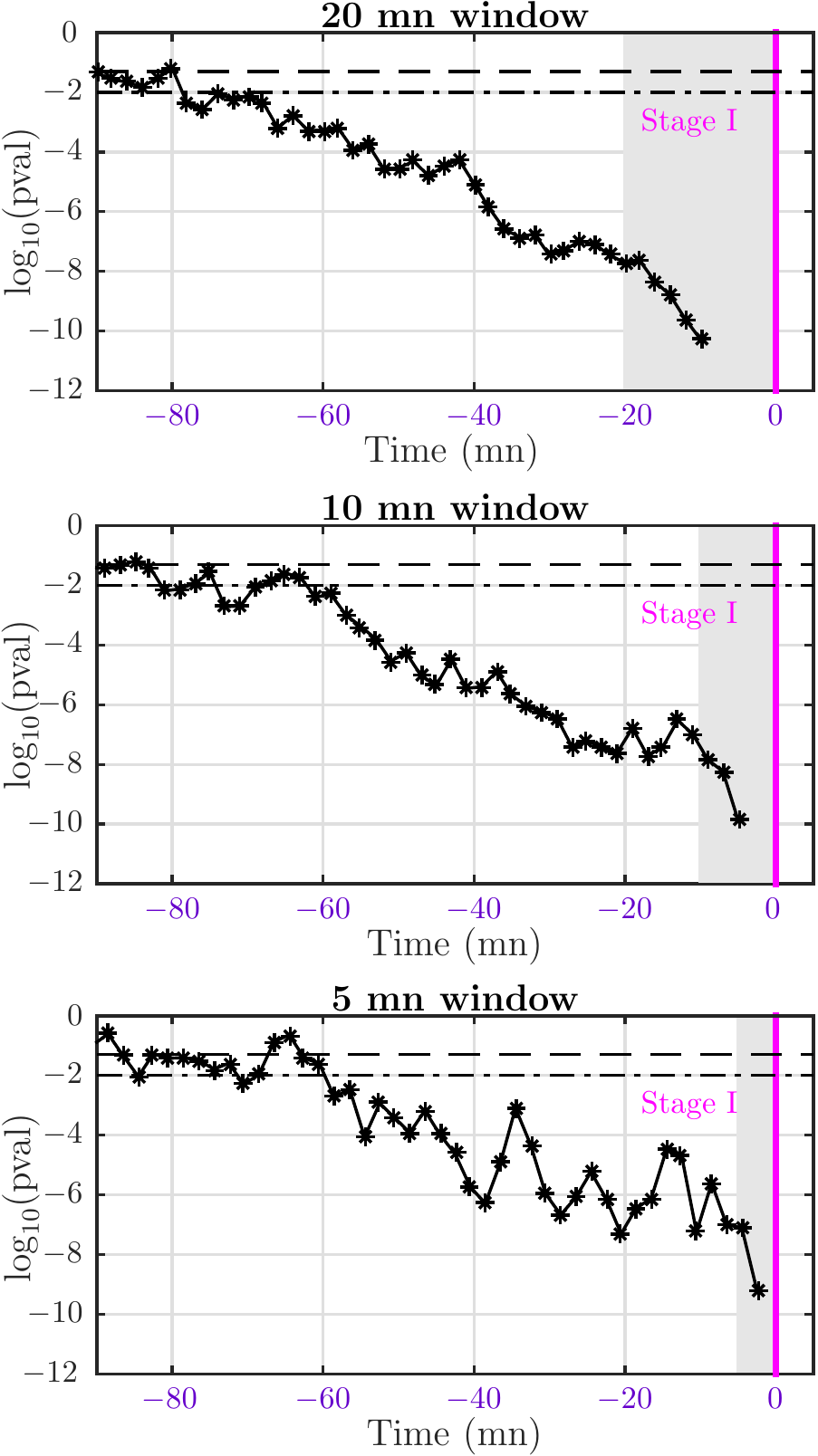}\\
\caption{Left: average AMI $\hat{I}^{(3,3)}$ for normal (black) and abnormal (red) subjects in dataset I (delivery occuring less then 15mn after pushing started). 
AMI is computed in windows of size $T=20$mn (first line), 10mn (second line) and 5mn (third line) shifted by 2mn.
Vertical magenta line indicate the beginning of stage 2 (pushing).
Right: corresponding $p$-value.
A single black + symbol in the AMI plot indicates a $p$-value lower than 0.05, two ++ indicate a $p$-value lower than 0.01.
\ch{The gray-shaded region represents the time window used to compute the last value of AMI.}
}
\label{fig:windows:groupI}
\end{center}
\end{figure}

In this first section, we focus on dataset I and probe the stage 1, including early labor, active labor, and transition.
Using the time at which stage 1 ends as a reference (setting it at $t_0=0$), we compute for each subject $\hat{I}^{(3,3)}$
in a set of time windows $[t_i-T, t_i]$, $0\le i\le 50$ of fixed size $T$ ending at $t_i=t_0-i*2$mn, so separated by 2mn. 
We perform this analysis for three window sizes $T \in[20, 10, 5]$mn.
The value of AMI computed in the $i$-th window is then assigned to time $t_i-T/2$, at the center of the interval.
By construction of dataset I, delivery occurs less than 15 minutes after pushing started, and can be as short as 1mn, so we completely discard data from stage 2. 
We then average the values of AMI over the population of normal subjects, and over the population of acidotic subjects, respectively.
Results, including $p$-values, are presented in Fig.\ref{fig:windows:groupI}.

A first observation is that AMI is always larger for acidotic subjects than for normal subjects.
As labor progresses, AMI increases in both populations, but the increase is stronger for acidotic subjects.
As a consequence, the $p$-value of the test decreases clearly, so the feature performs better and better when approaching delivery.
Detection of acidosis using the AMI feature and $T=20$mn can be obtained in dataset I as early as 80 minutes 
before entering in second stage. Using shorter windows, $T=10$mn or 5mn, detection is still reliable as early as one hour before stage 2. We interpret this reduced forecast of acidosis detection in dataset I as a direct consequence of the reduction of the statistics when the window size $T$ is reduced.

\subsubsection{Dataset II : delivery after pushing more than 15mn} 

\begin{figure}
\begin{center}
\includegraphics[width=0.45\textwidth]{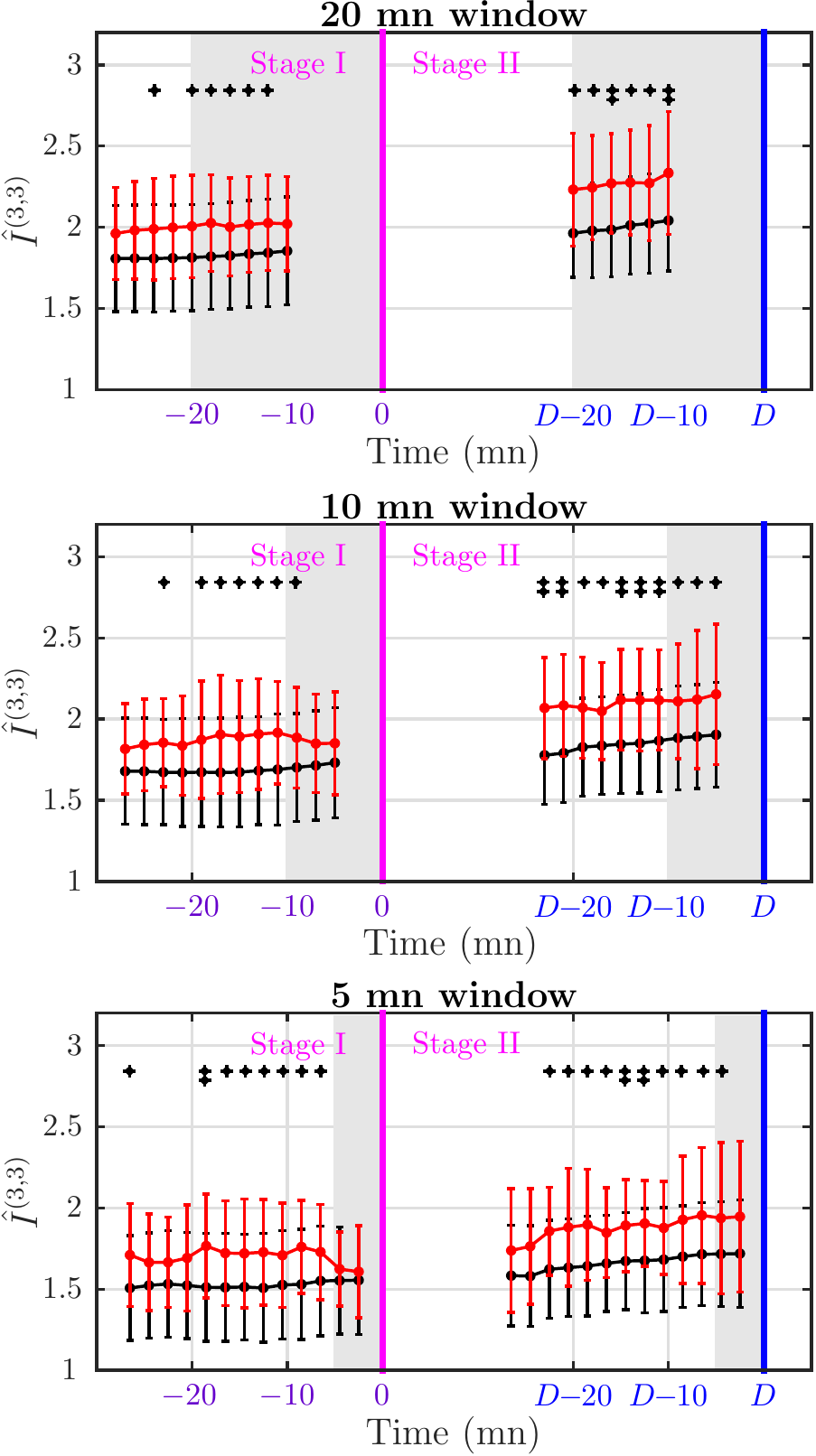}
\includegraphics[width=0.45\textwidth]{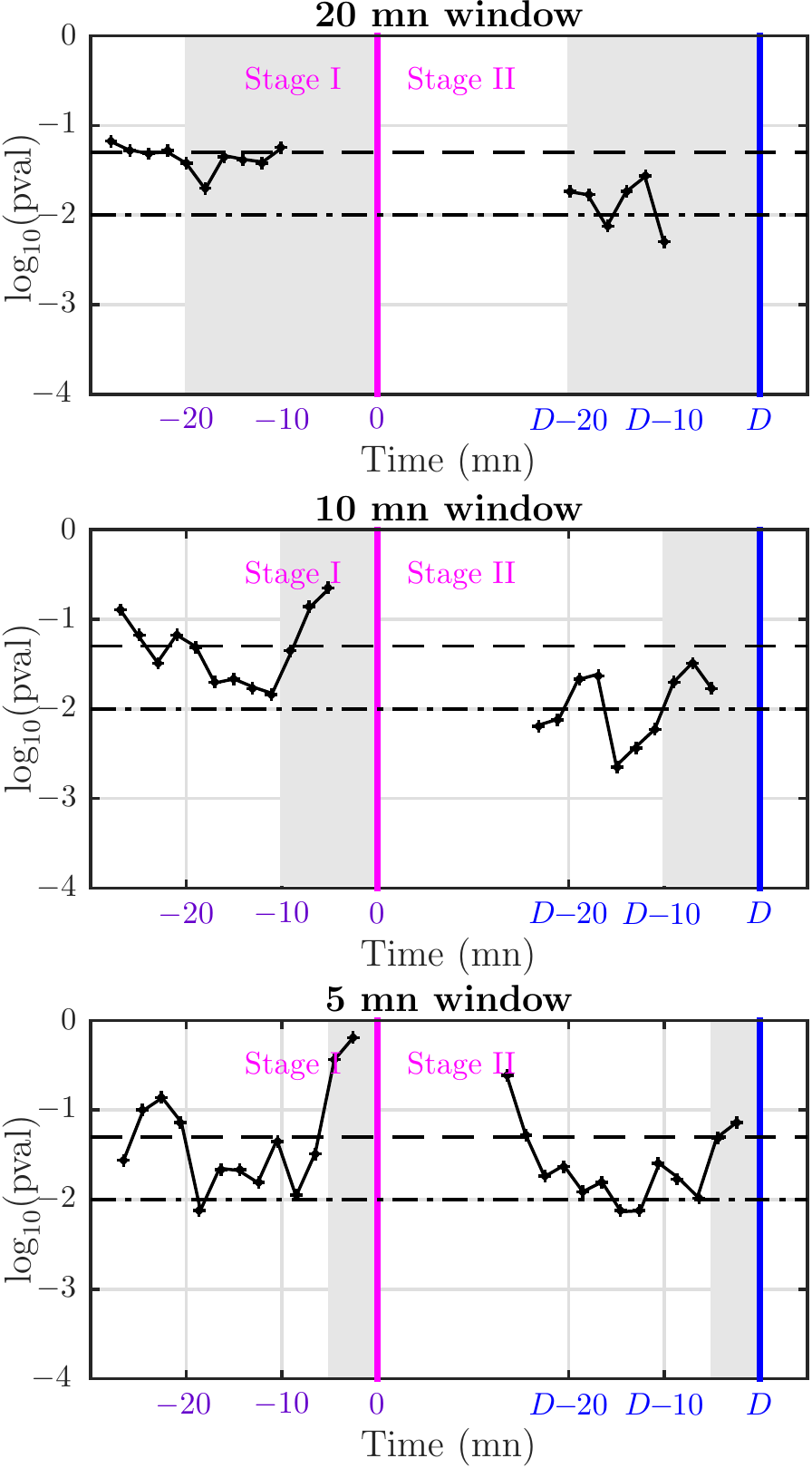}\\
\caption{Left: average AMI $\hat{I}^{(3,3)}$ for normal (black) and abnormal (red) subjects in dataset II (delivery occuring more then 15mn after pushing started). 
AMI is computed in windows of size $T=20$mn (first line), 10mn (second line) and 5mn (third line) shifted by 2mn.
Vertical magenta line indicate the beginning of stage 2 (pushing) and vertical blue line indicate delivery.
Right: corresponding $p$-value. 
Black + symbols in the AMI plot indicate a $p$-value lower than 5\% (+) or lower than 1\% (++).
\ch{The gray-shaded region represents the time window used to compute the last value of AMI.}
}
\label{fig:windows:groupII}
\end{center}
\end{figure}

For dataset II, we performed the same dynamical analysis as in the previous section, using the end of stage 1 as the reference time ($t_0=0$). 
Because there is now enough data in the pushing stage, we also perform the analysis of this stage using the delivery time ($t_0=D$) as the reference. 
All results are presented in Fig.~\ref{fig:windows:groupII}.
At the end of stage 1, we observe again that AMI is larger for acidodic subjects than for normal ones, but the difference is not significant in this group (see the corresponding $p$-value on the right of Fig.~\ref{fig:windows:groupII}).
The situation is identical at the end of stage 2, although we obtain a lower $p$-value in some windows.
The $p$-value does not decrease clearly when approaching delivery time, as it was in dataset I, see Fig.~\ref{fig:windows:groupI}.
For subjects in dataset II it is very difficult to make an early detection of acidosis.
However, we observe in Fig.~\ref{fig:windows:groupII} that the average AMI is significantly larger at the end of stage 2 than at the end of stage 1. The increase of AMI is larger for abnormal subjects.

\begin{figure}
\begin{center}
\includegraphics[width=0.45\textwidth]{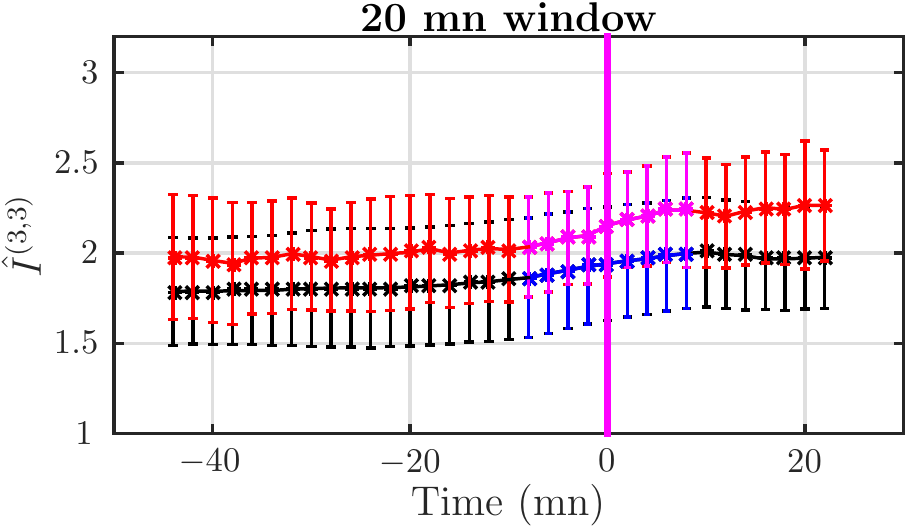}
\includegraphics[width=0.45\textwidth]{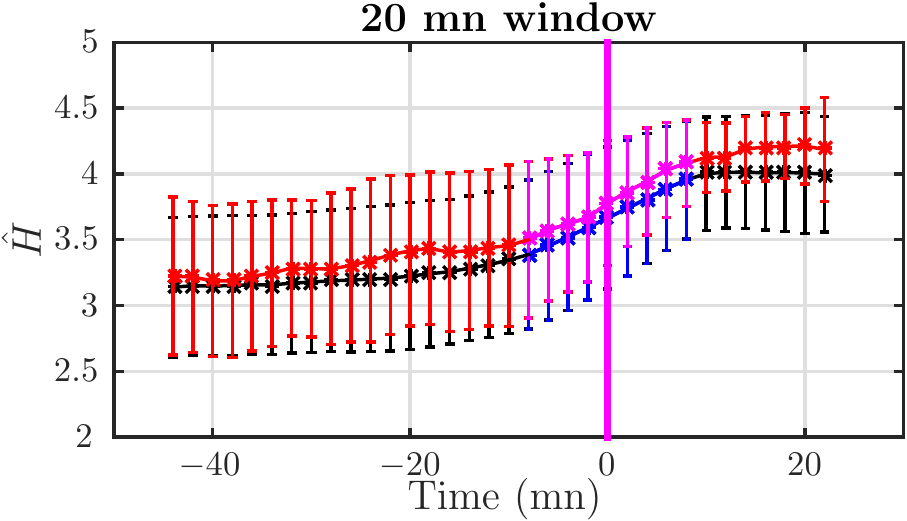}
\caption{Average behavior of AMI $\hat{I}^{(3,3)}$ (left) and Shannon entropy $\hat{H}$ (right) for normal (black) and acidotic (red) subjects in dataset II (delivery occuring more than 15mn after pushing started).
Quantities are computed in windows of size $T=20$mn.
Vertical magenta line indicate the beginning of stage 2 (pushing).
We have used a different color code for windows spanning both stages I and II: blue for normal subjects and 
magenta for acidotic ones.
}
\label{fig:windows:transient}
\end{center}
\end{figure}

To examine more precisely the dynamical increase of AMI, especially when entering stage 2, we computed $\hat{I}^{(3,3)}$ over an ensemble of windows of size $T=20$mn spanning continuously a large time interval that includes the end of the active labor and the beginning of the pushing stage. Results are reported in Fig.~\ref{fig:windows:transient}.
We see a continuous increase of AMI values when evolving from stage 1 to stage 2.
The increase is more important for abnormal subjects which corroborates the findings in Fig.~\ref{fig:windows:groupII}.
For smaller window sizes, the situation is less clear.

We also studied the dynamical evolution of the Shanon entropy estimate $\hat{H}$, which, together with the AMI, combines into the Shannon entropy rate (see eq.(\ref{eq:def:h:MI})). Changes in the Shannon entropy $H$ indicate changes in the probability density of the signal. 
Results are presented in Fig.~\ref{fig:windows:transient}, side by side with the AMI.
We observe a dramatic rise of the value of $\hat{H}$ when subjects evolve from stage 1 to stage 2. This increase is clearly observed for normal and abnormal subjects. 
No significant difference between normal and acidotic subjects is observed for this static quantity. 
The start of pushing implies a strong deformation of the probability density of the FHR, indicating strong perturbations of the FHR, for both normal and acidotic subjects.

\section{Discussion, conclusions and perspectives}

We now discuss interpretation of Shannon entropy and AMI measurements in different stages of labour. 
The fetuses are classified as normal or acidotic depending on a post-delivery pH measurement, which gives a diagnosis  of acidosis at delivery only. There is no information on the health of the fetuses during labor.

\ch{The physiological interpretation of a feature, and especially its relation to specific FHR patterns, e.g., like those detailed in~\cite{Freeman2002,Ayres-de-Campos2015}, is a difficult task that is only scarcely reported in the litterature~\cite{Signorini2003,Goncalves2008}.}
\ch{In this article, we have averaged our results over large numbers of (normal or acidodic) subjects, which jeopardizes any precise interpretation in terms of a specific FHR pattern that may appear only intermittently.}

\subsection{Acidosis detection in first stage}

\ch{We can nevertheless suggest that the value of the Shannon entropy $H$ is related to the frequency of decelerations in the FHR signals. Indeed, Shannon entropy strongly depends on the standard deviation of the signal (e.g., see eq.(\ref{eq:H:Gaussian})), which in turn depends on the variability in the observation window. A larger number of decelerations in the observation window deforms the PDF of the FHR signal by increasing its lower tail; in particular, this increases the width of the PDF, and hence increases the standard deviation and the Shannon entropy.
This explains our findings in Fig.~\ref{fig:boxplot20105} (for dataset I).}

When acidosis develops in first stage of labor, the Shannon auto mutual information estimator $\hat{I}^{(m,p)}$ significantly outperforms all other quantities both in terms of $p$-value and AUC. 
The performance of AMI is robust when tuning either the size of the observation --- and hence the number of points in the data~--- and the embedding dimensions $(m,p)$.
In addition, the performance slightly increases when the total embedding dimension $m+p$ increases; although one has to care about the curse of dimensionality.

For abnormal subjects from dataset II, AMI is not able to detect acidosis using data from stage 1.
This suggests that acidosis develops later, in the second stage of labor.

For all datasets, AMI \ch{computed with $\tau=0.5$s} is always larger for acidotic subjects than for normal subjects. This is in agreement with results obtained with ApEn and SampEn, which are both lower for acidotic subjects. This shows that FHR classified as abnormal \ch{have a stronger dependence structure at small scale} than normal ones. 
\ch{We can relate this increase of the dependence structure of acidotic FHR
to the short term variability and to its coupling with particular large scale patterns.
For example, a sinusoidal FHR pattern~\cite{Ayres-de-Campos2015}, especially if its duration is long, should give a larger value of the AMI, because its large scale dynamics is highly predictable. 
As another example, we expect {\em variable decelerations} (with an asymmetrical V-shape) and {\em late decelerations} (with a symmetrical U-shape and/or reduced variability) to impact AMI differently.
Of course, the choice of the embedding parameter $\tau$ is then crucial, and this is currently under investigation.}

AMI and entropy rates depend on the dynamics as they operate on time-embedded vectors.
AMI focuses on nonlinear temporal dynamics, while being insensitive to the dominant static information. 
AMI is independent of the standard deviation which on the contrary contributes strongly to the Shannon entropy.
This explains why AMI performs better than entropy rates estimates, such as ApEn$^{(m)}$, SampEn$^{(m)}$ and 
$\hat{h}^{(m)}$, which depend also on the standard deviation.

\subsection{Acidosis detection in second stage}

The results reported for stage 2 show a severe decrease in performance of the five estimated quantities.
Analysing stage 2 is far more challenging than analysing stage 1, which suggests that temporal dynamics in stage 2 differ notably for those of stage 1~ \cite{Spilka2016BSI} 
\ch{or simply that our database does not contain enough acidotic subject in that case}.
$\hat{I}^{(m,p)}$ achieves the best performance in terms of p-value and AUC;
this clearly underlines that the analysis of nonlinear temporal dynamics is critical for fetal acidosis detection in stage 2.
As in stage 1, the AMI is always larger in stage 2 for acidotic subjects than for normal subjects. 

\ch{Although the Shannon entropy computed from the last 20mn of stage 2 before delivery does not show a clear tendency in Fig.~\ref{fig:boxplot20105} for dataset II, looking at Figure~\ref{fig:windows:transient} clearly shows that $\hat{H}$ increases as labor progresses: this is probably related to the average increase of the number of decelerations, which is expected in both the normal and acidotic population.}

SampEn$^{(2)}$ is also able to perform discrimination in stage 2.
From these observations, one may envision the definition of a new estimator that would measure the auto mutual information using the R\'enyi order-2 entropy by applying eq.(\ref{eq:def:MI:h}). Nevertheless, it should be emphasized that R\'enyi order-$q$ entropy is lacking chain rule of conditional probabilities as soon as $q\neq 1$, which may jeopardize any practical use of such an estimator.

\subsection{Probing the dynamics}
 
Increasing the total embedding dimensions in AMI improves the performance in the detection of acidotic subjects, in both first and second stages. 
The best performance is found for different total embedding dimension in the two datasets. This suggests that FHR dynamics is different in each stage.

As seen in eq.~(\ref{eq:def:h:MI}), the Shannon entropy rate can be split in two contributions: one that depends only on static properties (the Shannon entropy, estimated by $\hat{H}$) and one that involves the signal dynamics (the auto mutual information, estimated by $\hat{I}^{(m,1)}$). By following the time evolution of these two parts, we were able to relate Shannon entropy $\hat{H}$ with the evolution of the labor and AMI not only with the evolution of the labor, but also with possible acidosis. 
Looking at subjects for which the pushing phase is longer than 15mn, it clearly appears that all fetuses are affected by the pushing, as evidenced by a large increase of the Shannon entropy $\hat{H}$, and a small increase of AMI.
Additionally, the increase of AMI is steeper for abnormal subjects than for normal ones, 
which may indicate different reactions to the pushing \ch{and can be related to specific pathological FHR patterns}. 
When the pushing stage is long (dataset II), fetuses reported as acidotic are not showing any sign of acidosis until prolongated pushing. These fetuses appear as normal until delivery is near.

When acidosis develops during the first stage of labor, in dataset I, we observe clearly that while AMI increases steadily till delivery for healthy fetuses, it increases faster for acidotic ones.
This suggests that acidotic fetuses in dataset I react to early labor, as early as one hour before pushing starts.
\ch{This could not only indicate that some fetuses are prone to acidosis, but also} may pave the way for an early detection of acidosis in this case.

\paragraph{Acknowledgments} This work was supported by the LABEX iMUST (ANR-10-LABX-0064) of Universit\'e de Lyon, within the program "Investissements d'Avenir" (ANR-11-IDEX-0007) operated by the French National Research Agency (ANR).

\end{document}